%% file: main.tex
\documentclass[sigconf]{acmart}

\AtBeginDocument{%
  \providecommand\BibTeX{{%
    \normalfont B\kern-0.5em{\scshape i\kern-0.25em b}\kern-0.8em\TeX}}}

\setcopyright{acmcopyright}
\copyrightyear{2021}
\acmYear{2021}
\acmConference[WWW]{}{April 25-29, 2022}{The Web Conference 2022}

\usepackage{subcaption}

\newcommand\eat[1]{}
\begin{document}

\title{PhishChain: A Decentralized and Transparent System to Blacklist Phishing URLs}

\author{Shehan Edirimannage\textsuperscript{†},  
  Mohamed Nabeel\textsuperscript{‡}, 
  Charith Elvitigala\textsuperscript{†},
  Chamath Keppitiyagama\textsuperscript{†}} 
\affiliation{ 
  \institution{\textsuperscript{†}University of Colombo School of Computing} 
  \institution{\textsuperscript{‡}Qatar Computing Research Institute}
  \country{Sri Lanka, Qatar}
}









\renewcommand{\shortauthors}{Edirimannage et al.}


\begin{abstract}
\input{abstract}
\end{abstract}


\maketitle

\section{Introduction}
\input{intro}


\section{System Architecture}\label{sec:archi}
\input{system}

\section{Demonstration}
\input{demo}

\section{Discussion}
\input{discussion}

\section{Conclusions}
\input{conclusion}


\end{document}

%% file: abstract.tex
Blacklists are a widely-used Internet security mechanism to protect Internet users from financial scams, malicious web pages and other cyber attacks based on blacklisted URLs. In this demo, we introduce PhishChain, a transparent and decentralized system to blacklisting phishing URLs. At present, public/private domain blacklists, such as PhishTank, CryptoScamDB, and APWG, are maintained by a centralized authority, but operate in a crowd sourcing fashion to create a manually verified blacklist periodically. In addition to being a single point of failure, the blacklisting process utilized by such systems is not transparent. We utilize the blockchain technology to support transparency and decentralization, where no single authority is controlling the blacklist and all operations are recorded in an immutable distributed ledger. Further, we design a page rank based truth discovery algorithm to assign a phishing score to each URL based on crowd sourced assessment of URLs. As an incentive for voluntary participation, we assign skill points to each user based on their participation in URL verification.


%% file: intro.tex
\begin{figure*}[!ht]
\centering
\begin{subfigure}{0.53\linewidth}
\frame{\includegraphics[width=\linewidth]{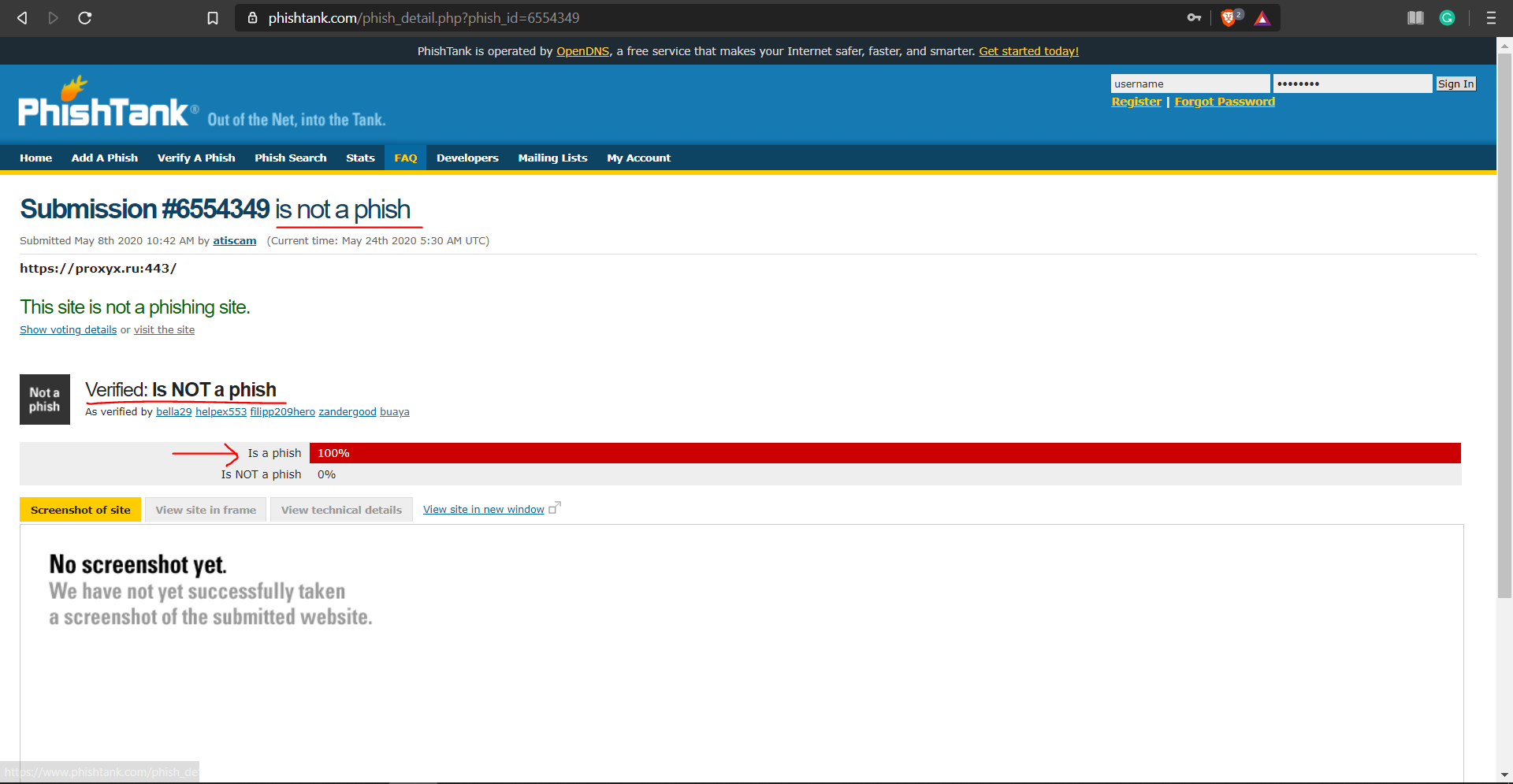}} 
\label{fig:case1}
\end{subfigure}
\hfill
\begin{subfigure}{0.36\linewidth}
\frame{\includegraphics[width=\linewidth]{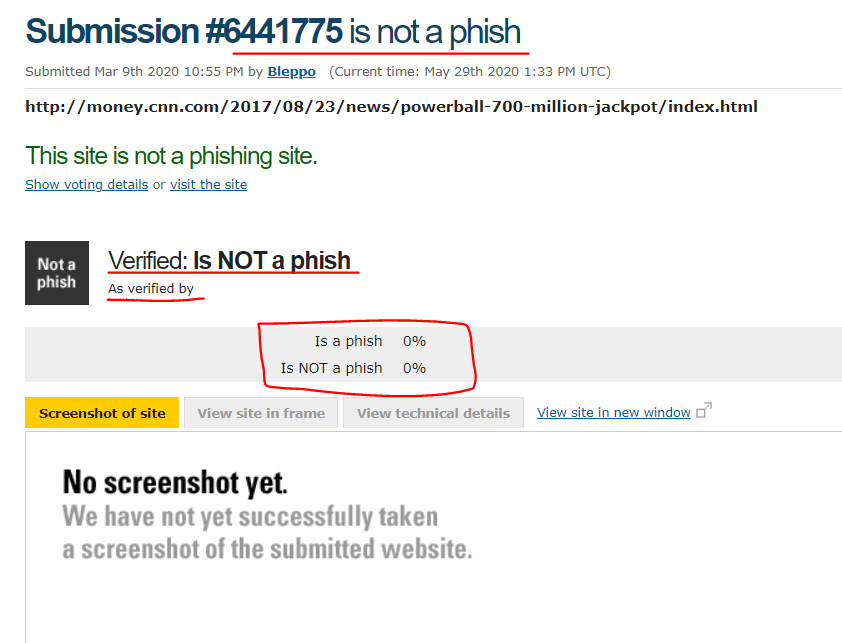}} 
\label{fig:case2}
\end{subfigure}
\caption{PhishTank inconsistent examples} 
\label{fig:inconsistent}
\end{figure*}

Crowdsourced phishing blacklists such as PhishTank, CryptoScamDB and APWG are quite useful in practice to defend against phishing attacks. Out of these services, PhishTank, has been the longest serving and most widely used crowd-sourced phishing blacklist. While it has been instrumental in maintaining a ledger of phishing URLs, it suffers from several limitations: (1) centralized architecture with a single point of failure, (2) lack of transparency, (3) lack of incentive to participate and (4) prone to manipulations. In fact, these are common issues for all existing centralized blacklisting systems. Focusing on PhishTank and for example, Figure~\ref{fig:inconsistent} shows inconsistent assignment of final labels to URLs in PhishTank. The figure on the left shows a URL marked as phishing even though all 5 crowd sourced verifiers ascertain it is phishing. The figure on the right, on the other hand, shows that a URL is marked as not phishing even though no one has verified it. It is not clear how PhishTank arrived at the final decision and raises concerns on the transparency.

We propose PhishChain, a transparent and decentralized system to blacklist phishing URLs in a crowd-sourced manner while addressing the limitations mentioned above. We envision a blacklisting system where a consortium of mostly targeted organizations such as Paypal, Apple, Microsoft and Facebook form an alliance to put our proposal to use. Our key design goals are as follows: (1) System should be decentralized where anyone can participate and no single authority should be controlling it, (2) All operations should be immutable and recorded, (3) URLs should be assigned a final phish score based on the participants' assessment, (4) Participants should be incentivized by assigning skill points instead of a cryptocurrency  in order to encourage voluntary participation~\footnote{Phishtank and APWG work through voluntary participation, but we observe that only a handful of participants contribute to the lists maintained.}, and (5) The system and data should be publicly available for anyone to consume.

There are a few notable efforts to build blockchain based decentralized solutions for phishing blacklisting or crowd sourcing in general. However, they suffer from one or two drawbacks. PhishLedger~\cite{phishledger} is a private blockchain solution for blacklisting phishing URLs, and they do not support crowd sourcing. Further, the system is not available for public consumption and is managed in a closed environment. Hence, the verifications performed in this system are questionable. CrowdBC~\cite{crowdbc} and ZebraLancer~\cite{zebralancer} are blockchain based crowd sourcing systems, but they rely on a cryptocurrency to maintain the functionality.

\textbf{Demonstration Scenarios}: PhishChain is designed as the next generation decentralized phishing blacklisting system. We show the following three use cases. (1) URL verification, (2) URL detailed dashobard and transaction, and (3) URL verification graph and timeline along with the phish score and verifier skill points.

%% file: system.tex
\subsection{Overall System Design}
\begin{figure*}[t]
\centering
    \includegraphics[width=0.8\textwidth]{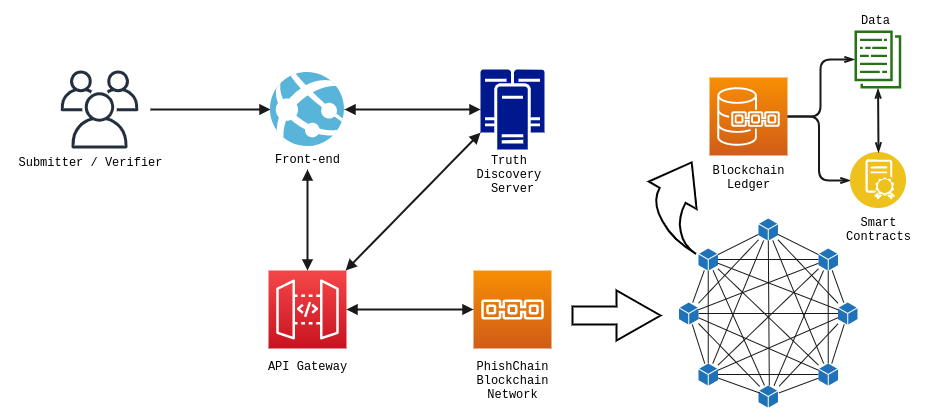}
\caption{High Level System Design}
\label{fig:design}
\vspace{-6mm}
\end{figure*}

Figure~\ref{fig:design} shows the overall system architecture. The key components in the system are blockchain network, smart contracts, truth discovery algorithm, and API services modules. The system supports three main functions in addition to user and system management functionality: (1) URL submission, (2) URL verification and (3) URL lookup. 

\textbf{Blockchain Network Module}:
Blockchain network is the backbone of the decentralized infrastructure. Each node in our blockchain network is either a validator or a normal node. A validator node validates each and every transaction whereas a normal node is a read-only participant. Note that the validation node only participate in the blockchain block verification process similar to committing data to the blockchain ledger, whereas the verifier participates in the URL verification process itself. Istanbul Byzantine Fault Tolerance is utilized as the consensus algorithm since our threat model assumes that there can be malicious participants. In order to meet our system goals, a consortium type blockchain system called, Quorum~\cite{quorum}, is employed as the blockchain network. Tessera~\cite{tessera} is used to manage transactions of the Quorum network.

\textbf{Smart Contract Module}:
The crowd source functionality is implemented using smart contracts (Solidity version 0.5.0). We categorize the smart contracts into three main groups: (1) user management, (2) URL management and (3) URL verification. 

\textbf{Truth Discovery Module}:
In our crowd sourcing environment, URL verifiers are the workers and the task is to assess if a URL is phishing or non phishing. Due to the varying degree of expertise and experience of verifiers, one cannot simply take, for example, the majority answer as the final phish score of a URL. Instead, a truth discovery algorithm which takes potentially conflicting status for a URL and verifier information, and arrives at the final phish score, should be employed. We observe that existing truth discovery algorithms developed by the database community~\cite{surveyon_td} are performing poorly on our problem as these algorithms assume that the majority of the verifiers respond to each URL verification task, which is not true for verifying URLs. Our retrospective analysis of PhishTank URLs show that only a handful of verifiers verify a given URL even though there are thousands of them altogether. Hence, we design a new truth discovery algorithm by leveraging page ranking (PR) algorithm on \emph{verifier-verifier graph}. We construct the verifier-verifier graph by using the ``\emph{follower}'' relationship. We define verifier $v_1$ follows $v_2$ for a given URL if $v_1$ verifies the URL before $v_2$. A directed edge from $v_1$ to $v_2$ is created in this case. The generated graph represent real world relationships among users which captures how users are following others and how they are being followed. Applying the page rank algorithm on top of this graph computes a rank, a real value number between 0 and 1, for each node (i.e. user), indicating how important it is with respect to the whole user population. It follows the intuition that, the more the number of followers are, the more recognition a user receive from its peers. 
We utilize the computed ranks to calculate the weighted voting score for each URL and assess if it is phishing or not.

Table~\ref{tab:perf} shows how our algorithm outperforms in comparison to popular truth discovery algorithms on a real-world dataset obtained from PhishTank. We collect URLs submitted and verified in PhishTank from January 1$^{st}$ 2020 to December 23$^{rd}$ 2020. The cleaned dataset contains ~ 17,000 phishing URLs and ~ 6,000 non phishing URLs. We scrape the additional information of each of these URLs to identify the verifiers and their order of verification. We use a balanced dataset of 6,000 URLs from each class for the evaluation. 

We use our PR based algorithm to assign skill points to verifiers as well as final phish score to URLs. The skill points of verifiers are computed based on the verifier reputation score returned by the truth discovery algorithm and the proportion of correctly labeled URLs. The phish score of a URL is computed with the weighted score of phishing and non phishing verifiers.

Note that in PhishChain system, we enforce transparency and consistency on the data by following the consensus algorithm. In other words, all the data related to URLs (URLs themselves, votes, status) on the blockchain is consistent and hardened from unauthorized modifications. While our system has a separate truth discovery server for URL verification due to the practical limitations in the Solidity framework utilized, as the algorithm and the data are publicly available and verifiable, any user can verify the the results produced by the URL verification proces is indeed correct.


\begin{table}[!th]
\caption{Truth Discovery Algorithm Comparison} 
\label{tab:perf}
\centering
\begin{tabular}{| p{1.4in} || p{0.4in} | p{0.4in} | p{0.4in}|}
\hline
\textbf{Algorithm} & \textbf{Acc.} & \textbf{Prec.} & \textbf{Rec.}\\
\hline
\hline
 Our Approach & \textbf{95.45\%} & \textbf{96.74\%} & 94.31\%\\ 
 \hline
 EM  & 93.71\% & 91.01\% & \textbf{97.75\%}\\
 \hline
 GLAD & 93.98\% & 91.72\% & 97.39\%\\
 \hline
\end{tabular}
\end{table}

\textbf{API Services Module}:
This module facilitates the communication with the blockchain network. The demo web application we build is a consumer of this API.

%% file: demo.tex
\begin{figure}[t]
\centering
    \frame{\includegraphics[width=0.49\textwidth]{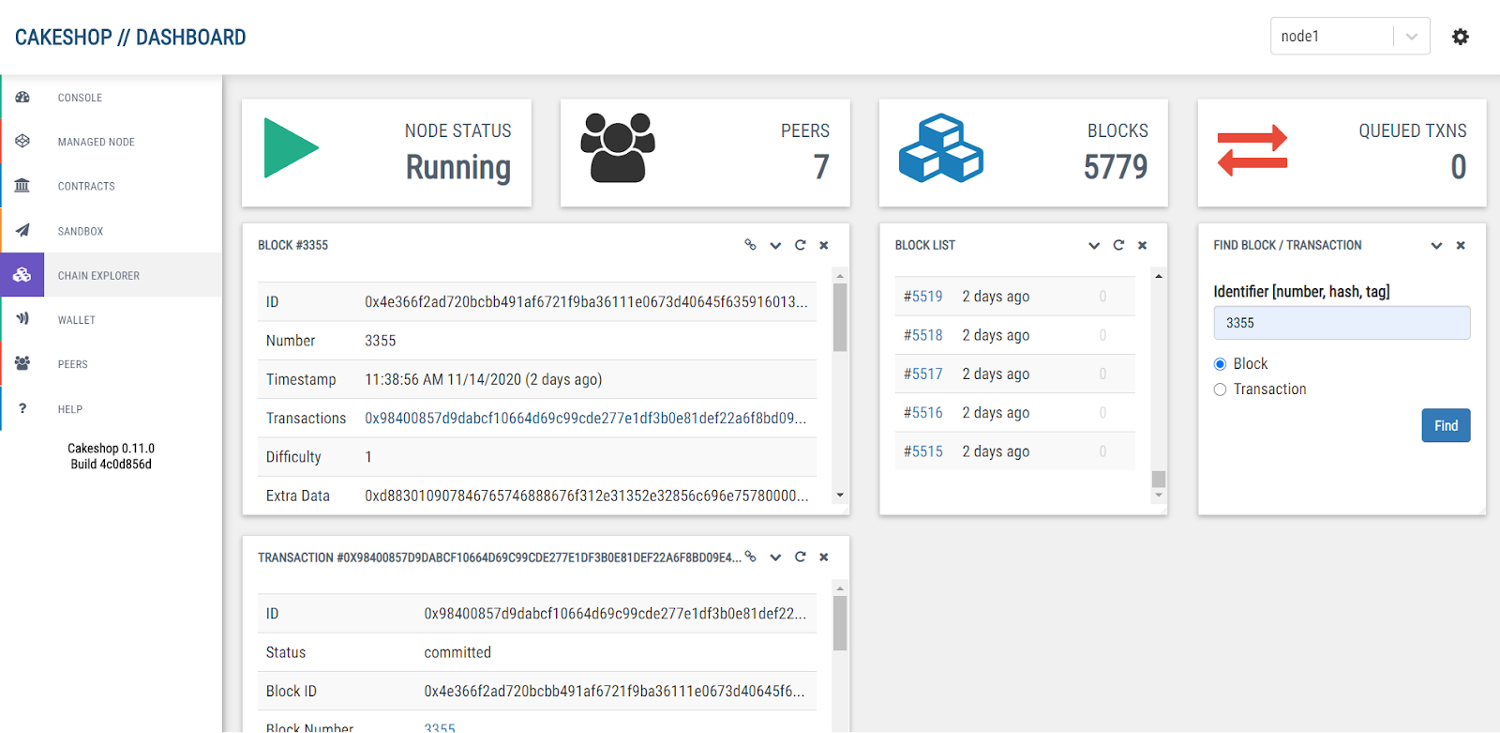}}
\caption{Cakeshop blockchain explorer showing the Quorum}
\label{fig:quorum}
\end{figure}

In this section, we describe the main use cases of our system. As shown in Figure~\ref{fig:quorum}, we setup a consortium blockchain consisting of 7 validation nodes each possessing a copy of the blockchain ledger and participating in validating each transaction. In a real-world setup, these validation nodes are to be hosted by organizations mostly targeted by phishers (e.g. Paypal, Facebook, Microsoft) as such a system is in their best interest to defend against phishing attacks and a collective effort is likely to yield a significantly better blocklist compared to having individual systems.

\subsection{URL Verification}
\begin{figure}[t]
\centering
    \frame{\includegraphics[width=0.49\textwidth]{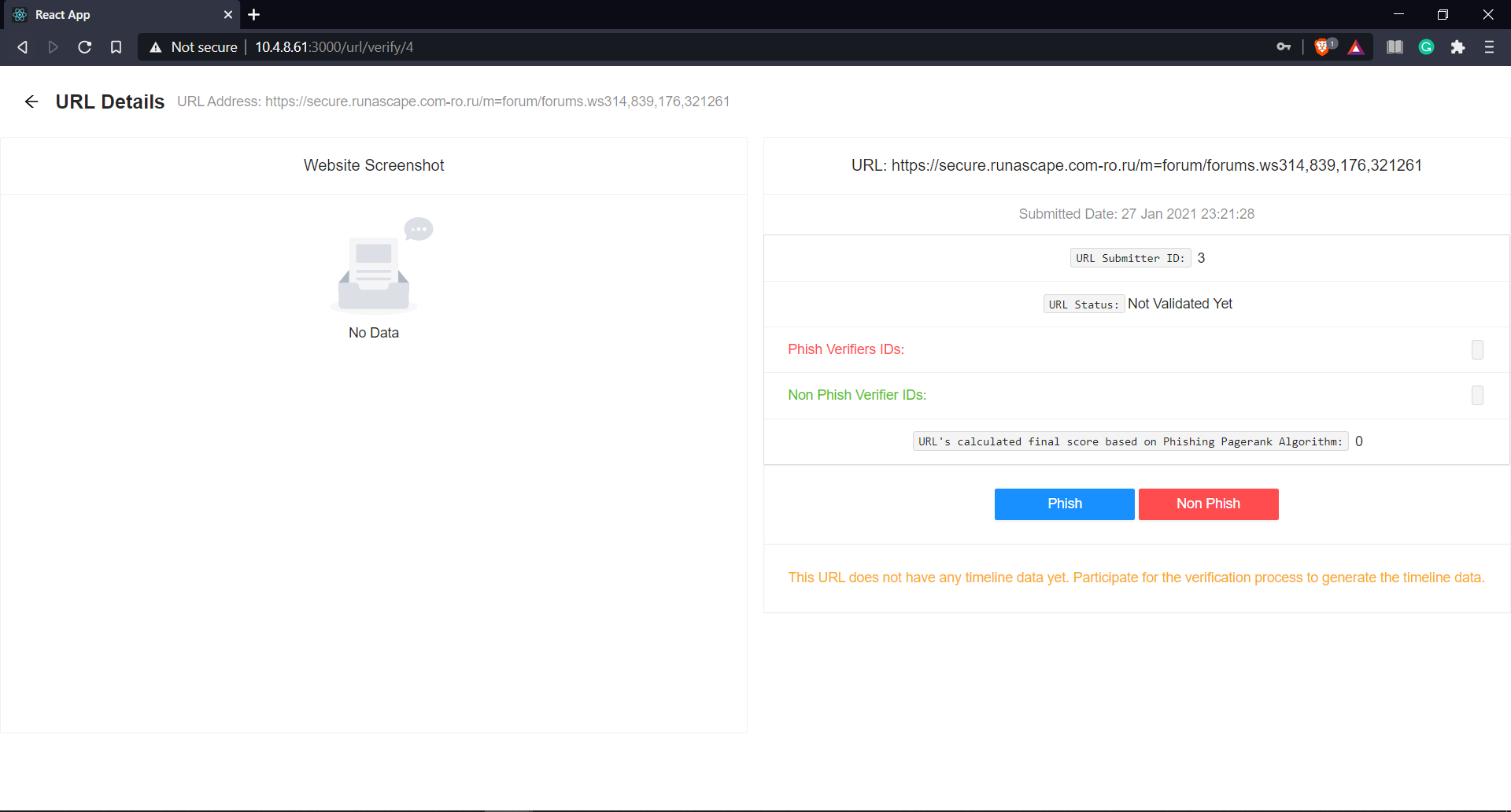}}
\caption{UI for URL Verification}
\label{fig:verify}
\end{figure}

\begin{figure}[t]
\centering
    \frame{\includegraphics[width=0.4\textwidth]{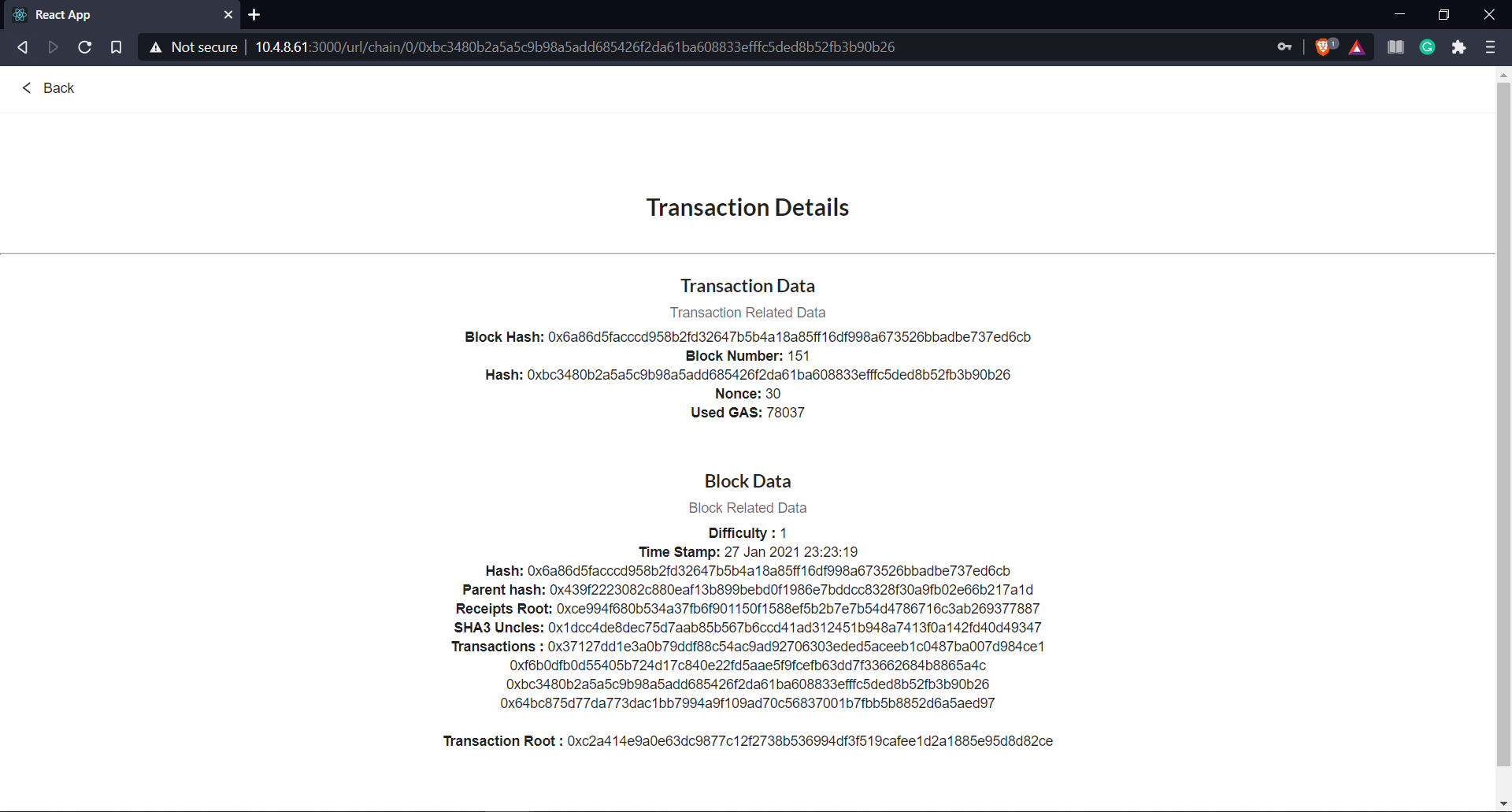}}
\caption{UI for URL Verification Transaction Details}
\label{fig:tx}
\end{figure}

Any user can sign up and participate in the process of submission and verification of URLs. Figure~\ref{fig:verify} shows the UI for the verification of a given URL as phishing or not phishing. Once user submits their verdict, the validation nodes validate the transaction adhering to the consensus algorithm resulting the transaction shown in Figure~\ref{fig:tx}. Once verified, if there are at least 3 votes for the URL, truth discovery model is invoked and assigns the ULR's calculated score where a positive score indicates that the URL is a phishing URL. Note that the users in here are real-world users with varying degree of domain expertise who may either submit URLs or verify whether the URL is phishing or not. Further, in order to limit the abuse of the system, PhishChain enforces that each URL submitted to the system should be accompanied with a valid email referring to the the URL~\footnote{Similar to existing blocklists, the goal of PhishChain to assess suspicious URLs found in likely phishing emails.}.

\subsection{URL Detailed Dashboard}
\begin{figure}[t]
\centering
    \frame{\includegraphics[width=0.4\textwidth]{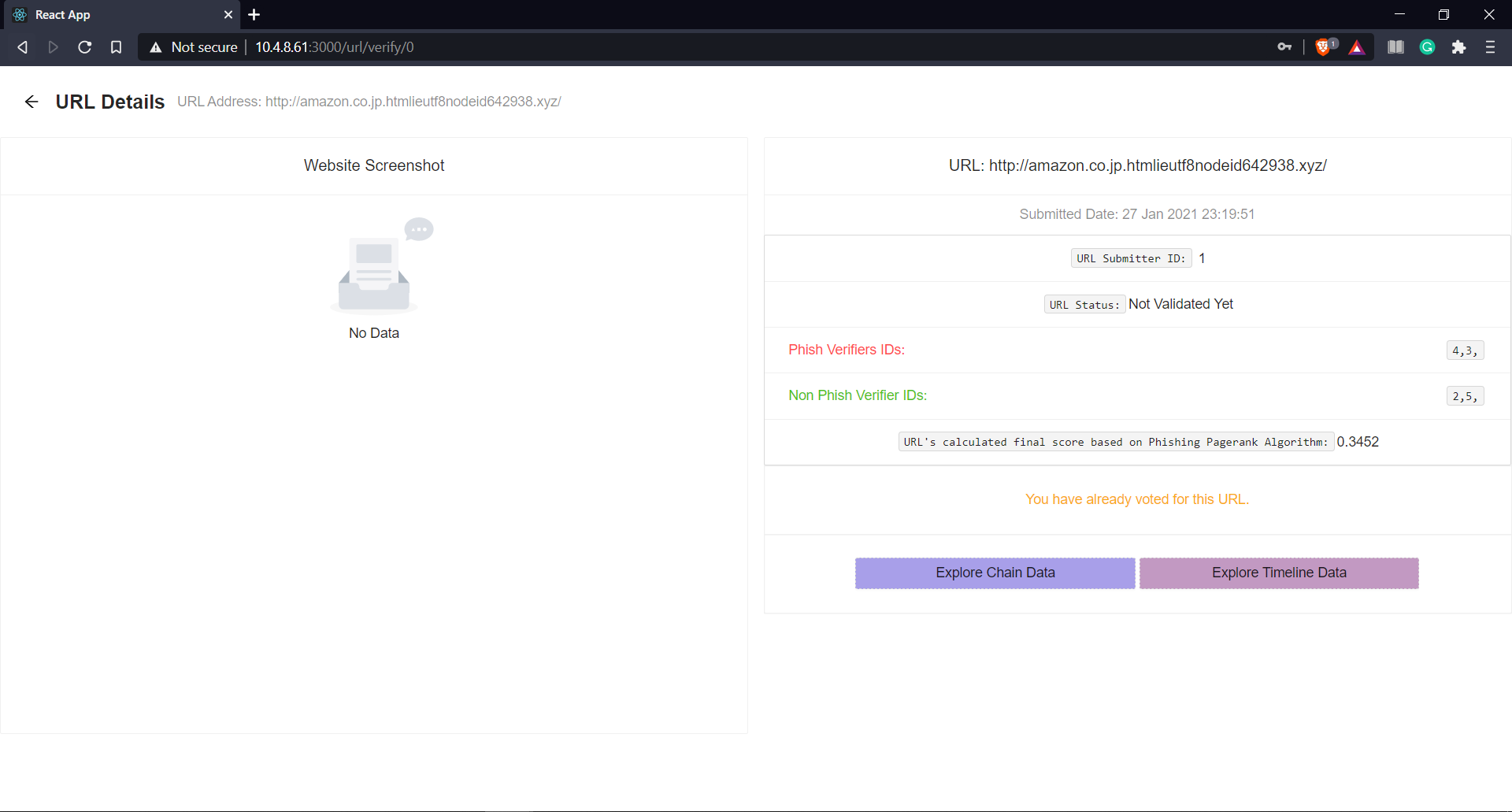}}
\caption{UI for URL Detailed Dashboard}
\label{fig:verified}
\end{figure}

Figure~\ref{fig:verified} shows the details of the URL submitted by user 1, and it is verified by users 3 and 4 as phishing and users 2 and 5 as non phishing. The system assigns an overall final phishing score of 0.3452 based on PR based truth discovery algorithm. Since it is a positive value, the system concludes that the URL is a phishing URL at that point in time.

\subsection{URL Verification Graph and Timeline}
\begin{figure}[t]
\centering
    \frame{\includegraphics[width=0.49\textwidth]{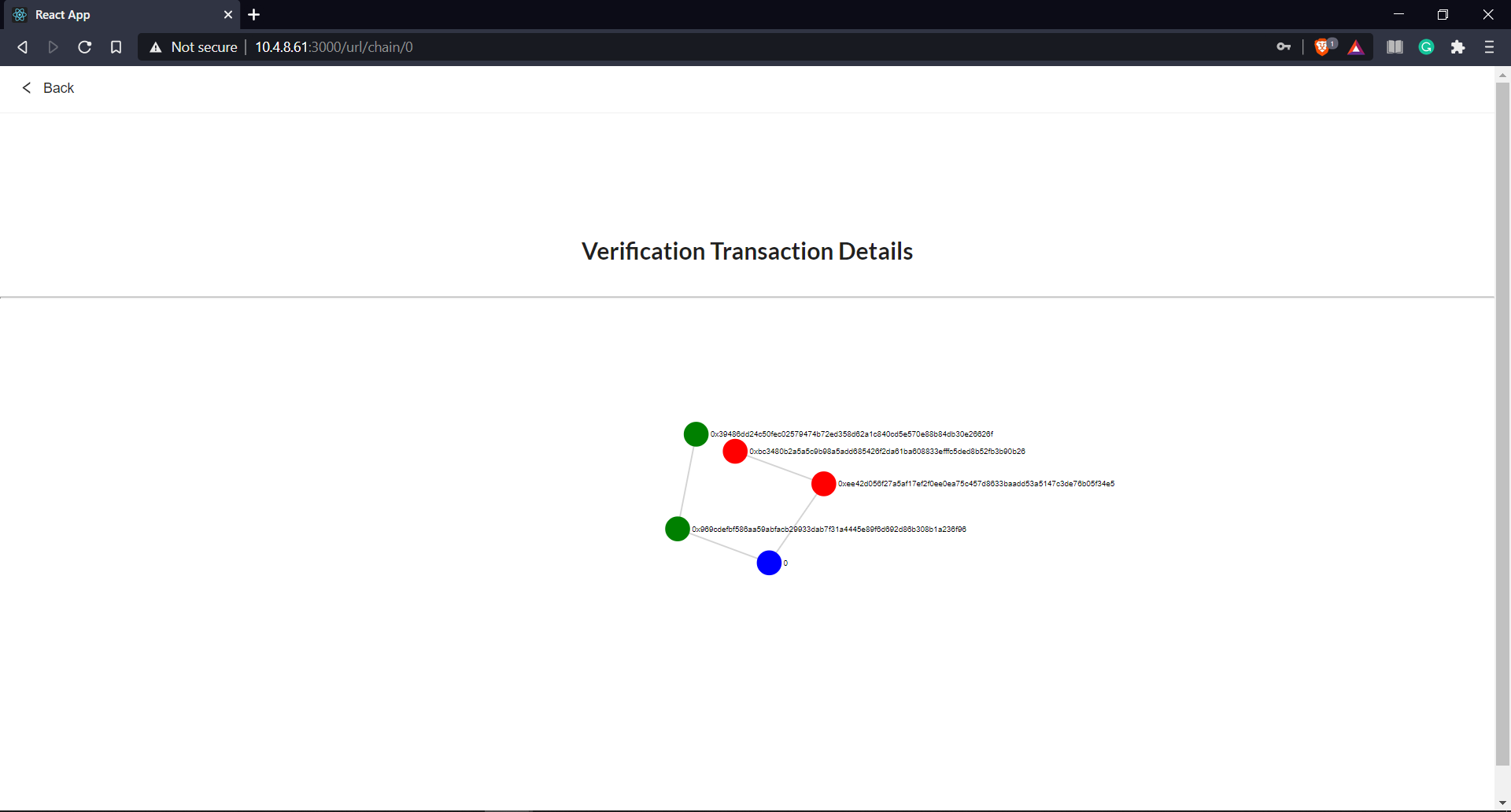}}
\caption{UI for URL Verification Graph}
\label{fig:vgraph}
\end{figure}

\begin{figure}[t]
\centering
    \frame{\includegraphics[width=0.49\textwidth]{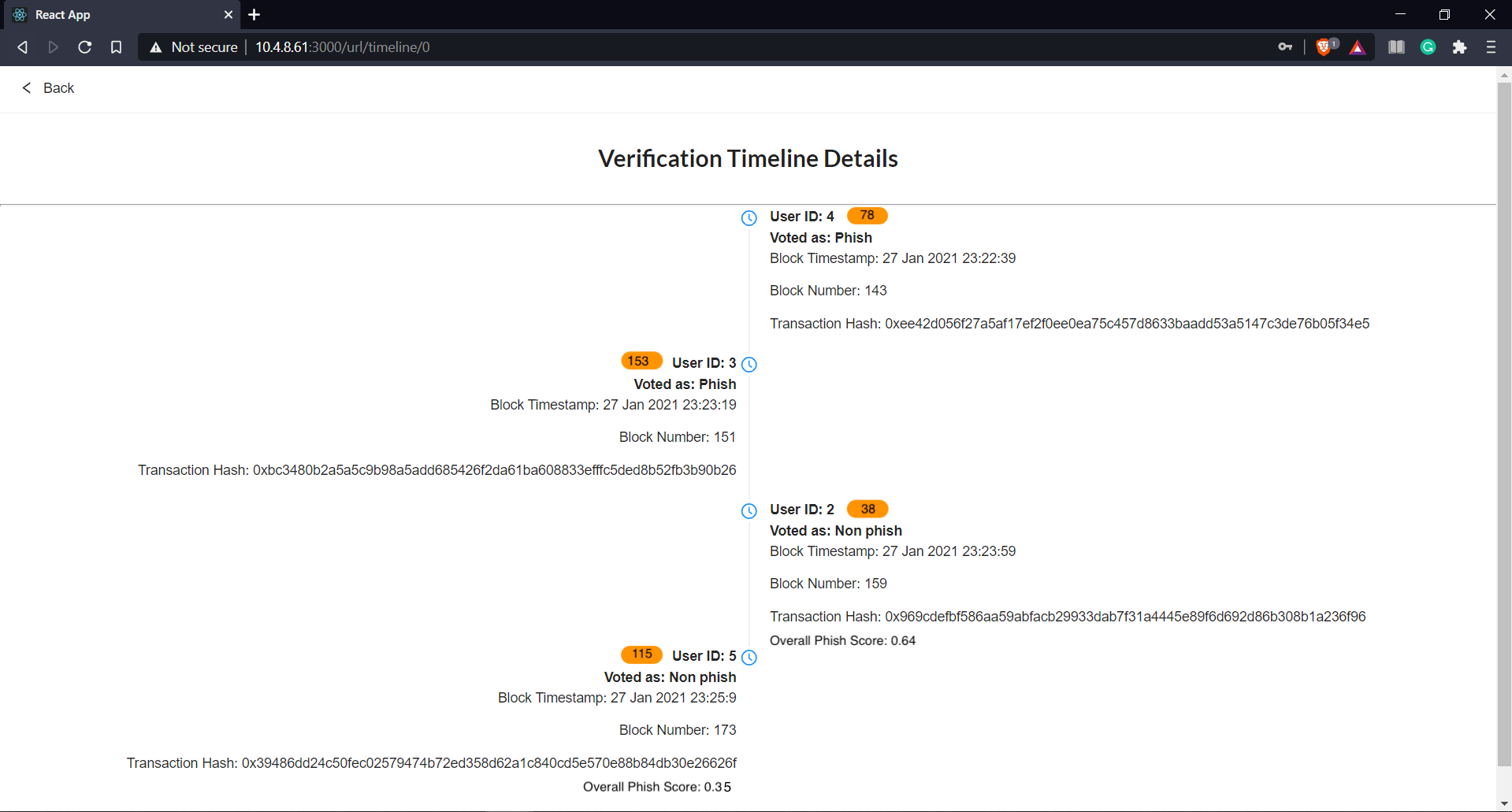}}
\caption{UI for URL Verification Timeline}
\label{fig:vtimeline}
\end{figure}

Figure~\ref{fig:vgraph} and Figure~\ref{fig:vtimeline} show the verification graph and the timeline of verification respectively. Based on the participation in the system and the proportion of accurately labeled URLs, each user is assigned skill points. For example, user 1 has the highest skill points of 153. Notice that after the 3$^{rd}$ vote, the URL is marked as phishing with a score of 0.64, but after the 4$^{th}$ vote, URL is still marked as phishing but with a lower phishing score of 0.35. The phishing score changes with every additional vote as truth discovery algorithm is invoked on affected URLs. Such a detailed timeline easily allows an analyst to not only assess the latest status but also understand how our system arrived at the decision.

%% file: discussion.tex

The motivation for introducing skill points is inspired from the question answering systems such as StackOverflow and StackExchange. Such systems show that even though participants do not receive any monetary benefits by contributing, these participants attain various ``expertise'' levels that have far more significance in their respective domains. We believe a similar system for malicious URL verification could assist solve the problem of the limited participation in the volunteer based blacklisting systems currently available.

In this work, we assume that users do not maliciously try to manipulate the truth discovery based system. However, one may extend PhishChain to defend against such manipulations by introducing additional mechanisms that the reputation/recommender system community has already made available~\cite{friedman2007manipulation}.

%% file: conclusion.tex
We propose, PhishChain, a truly transparent and decentralized phishing URL blacklisting system that is managed by a consortium instead of a centralized entity. Based on the proposed page rank based truth discovery algorithm, we compute a phish score for each URL as well as verifier skill points. Smart contracts are employed to implement the proposed functionality over the Quorum blockchain. We demonstrate three use cases: (1) URL verification, (2) URL dashboard and its transaction, and (3) URL verification graph and timeline. We believe our approach is generally applicable to any crowd sourced blacklisting system.